# On the predictability of the number of convex vertices

## Jean Souviron[1]

COGITECH Jean Souviron
613 D'Ailleboust
Montréal,  Quebec,
H2R 1K2 Canada

*JeanSouviron@hotmail.com*

**Abstract :**  Convex hulls are a fundamental geometric tool used in a number of algorithms. As a side-effect of exhaustive tests for an algorithm for which a convex hull computation was the first step, interesting experimental results were found and are the sunject of this paper. They establish that the number of convex vertices of natural datasets can be predicted, if not precisely at least within a defined range. Namely it was found that the number of convex vertices of a dataset of $N$ points lies in the range $2.35\ N^{0.091} \le h \le 19.19\ N^{0.091}$. This range obviously does not describe neither natural nor artificial worst-cases but corresponds to the distributions of natural data. This can be used for instance to define a starting size for pre-allocated arrays or to evaluate output-sensitive algorithms. A further consequence of these results is that the random models of data used to test convex hull algorithms should be bounded by rectangles and not as they usually are by circles if they want to represent accurately natural datasets.

---

[1] Dr Jean Souviron, Ph.D.1984, has been an independent consultant in scientific programming since 1994..

## 1. Introduction

Convex hulls are the stepping stone for a number of algorithms in computational geometry. The best algorithms to compute these convex hulls are *output-sensitive* algorithms, meaning that their computational complexity depends both of the input *and* output sizes. If $N$ is the number of points in the dataset and $h$ is the number of convex vertices, the most well-known of such algorithms for computing convex hulls are Kirkpatrick & Seidel[6]'s and Chan[4]'s, both in O($N \log h$). As part of the testing of an algorithm (*Souviron*[9]) on the subject of the *non*-convex contouring of a set of points, a detailed run on a number of datasets was performed. During these tests the author stumbled upon a very unexpected result: although at first thought one could not think of a relationship between $N$ and $h$, as intuitively a small dataset could have the same number of convex vertices than a very large one, a trend seemed to appear linking these two numbers. In order to check whether these results were maybe biased by the origin of the datasets further tests were made on datasets of various origins. This paper presents the results of this study.

## 2. The first results

The data used in the tests were lightning strike locations delimiting storm cells. They were obtained in two days in the summer of 1998 through the CLDN (*Canadian Lightning Detection Network*), courtesy of Environment Canada. Regrouped so that they encompassed a defined time period (*from 10 minutes up to 2 hours*) and a defined resolution, aiming at checking from storm cells to storm fronts (*from 2.5 up to 350 km minimum distance between locations*), they formed a sample of 629 datasets ranging from 4 to more than 93,000 points.

While displaying the number of convex vertices versus the number of dataset points using linear scales did not show any obvious trend linking both quantities, using log-log scales as shown in Figure 1 on the other hand seems to display a link between the number of convex vertices and the number of datatset points, which as above-mentioned was totally unexpected.

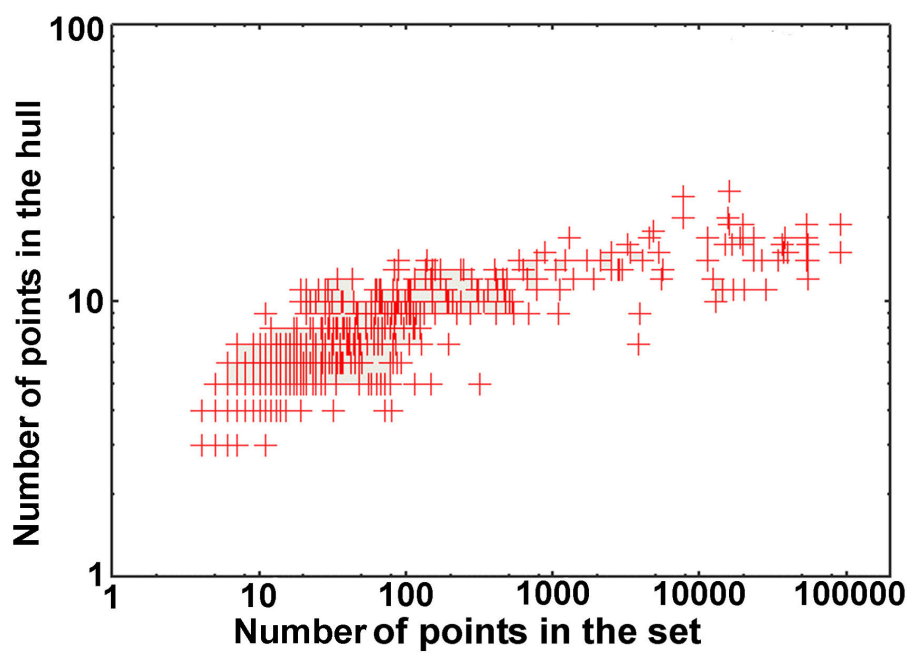

**Figure 1.** Number of convex vertices *vs* number of dataset points for lightning data

## 3. Extension to other data

In order to confirm or infirm this trend, as maybe an unknown factor depending on the lightning or storm phenomenon introduced a bias, and although there were more than 600 sets in the lightning datasets, extensive checks were needed using datasets of diverse origins and various sizes, but especially very large datasets. Subsets of public geo-political information files[1], medical images[2], subsets of some botanical data[3] and two geographical maps[4] containing up to 760,000 points were tested. Results were confirmed by other private datasets. Figure 2 shows the results for all datatsets. Although there is some noise, the trend is definitively there. Figure 3 shows the best range limits which can be derived from these data.

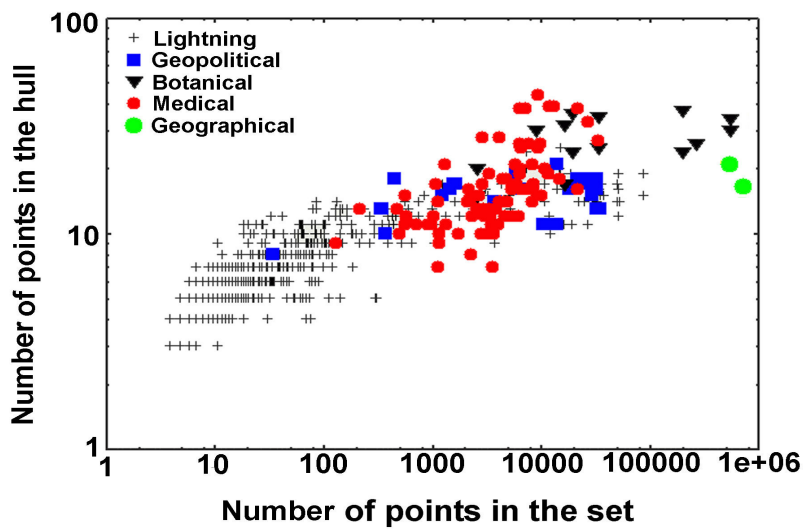

**Figure 2.** Same as Figure 1 but with all the data superimposed

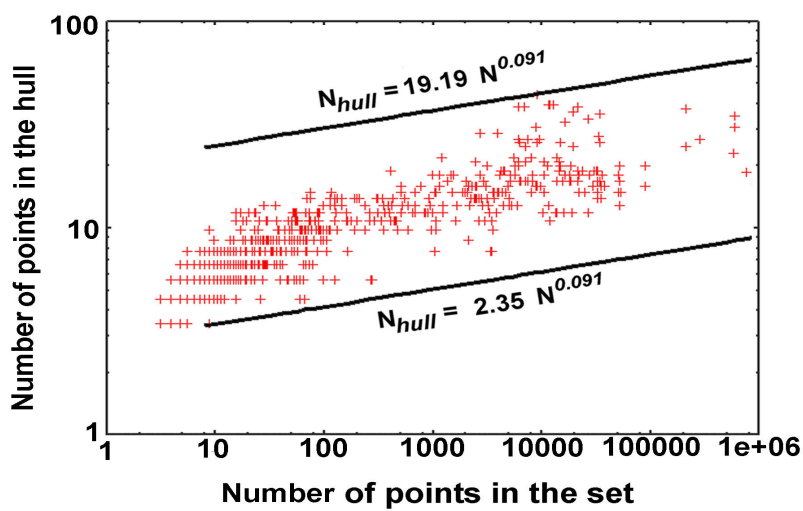

**Figure 3.** Fractional-power law defining the range of convex vertices *vs* the number of points
*Red crosses represent all the datasets (766). The N exponent is 0.091.*

---

[1] RGC dataset (*France's cities geographic directory*) of IGN (*french National Geographic Institute*). 31 files were obtained by selecting several population ranges as well as several city's area ranges.
[2] 10 grainy images of most of the categories of the 2D Hela databank of the US National Institute of Aging were thresholded to various high levels as to obtain 88 files of irregular and separated points.
[3] Cover dataset from the UCI Machine Learning Datasets Repository. 16 files were obtained by selecting the different cover types. (*extreme density*)
[4] High resolution (*down to 10-metres accuracy in some areas*) hydrological network and coastal map of North America courtesy of Environment Canada.

Finally, a last source of data was checked. These datasets are computer-generated examples of clusters used for research purposes[1]. The results for this particular source are coherent with previously found results, but for one dataset, as Figure 4 shows.

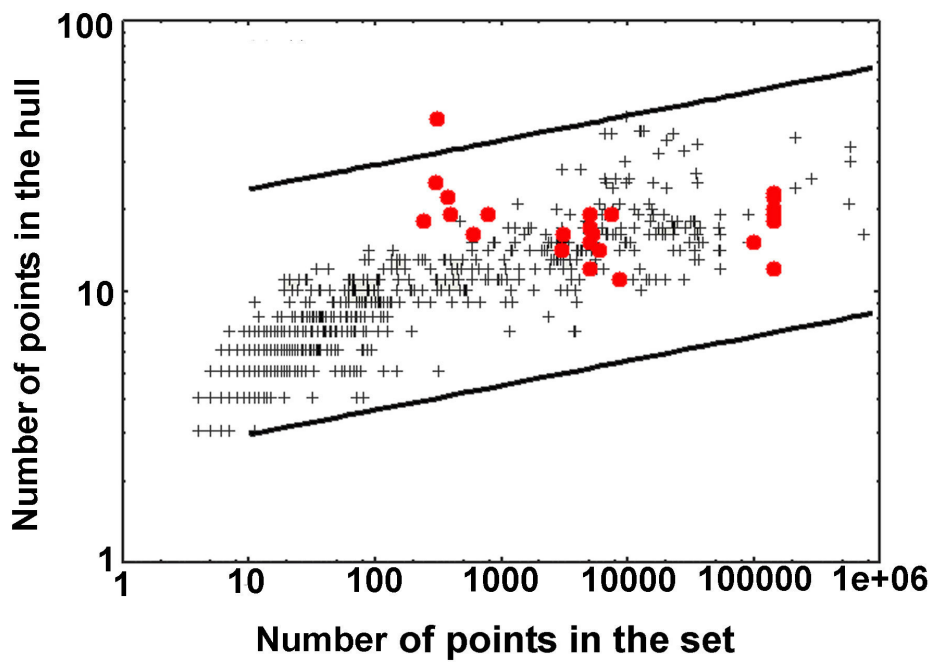

**Figure 4.** Same graph with the clusters datasets added

*Red dots are for the cluster datasets, black crosses represent all the other datasets*

As *only one* dataset falls outside the above-mentioned limits, this case was investigated and it appears that it is a special case, as demonstrated by Figure 5 below.

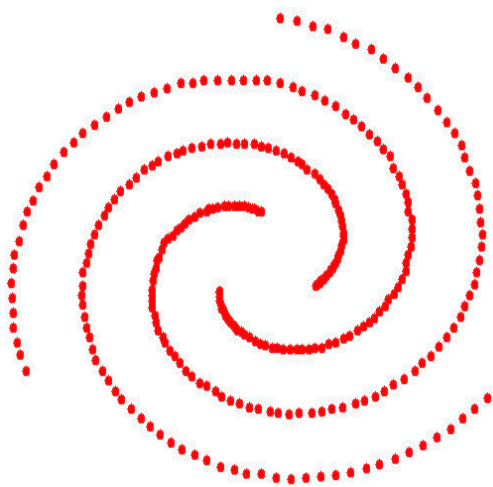

**Figure 5.** The special case of the clusters database

This figure is a partial worst-case type of example for convex hulls as a number of points lie on the outisde circle even if they do not *all* lie on the circle's perimetre. This counter-example prompted a further investigation on random sets of data.

## 4. Random sets of data and the representation of reality

Random sets of data are of widespread use as a test tool for a number of algorithm techniques and especially for convex hull computations. Two types of random data are usually used: circle-bound and rectangle-bound random data. Two kinds of circle-bound random data can be generated: either uniformly or centrally distributed. Examples of each are shown below.

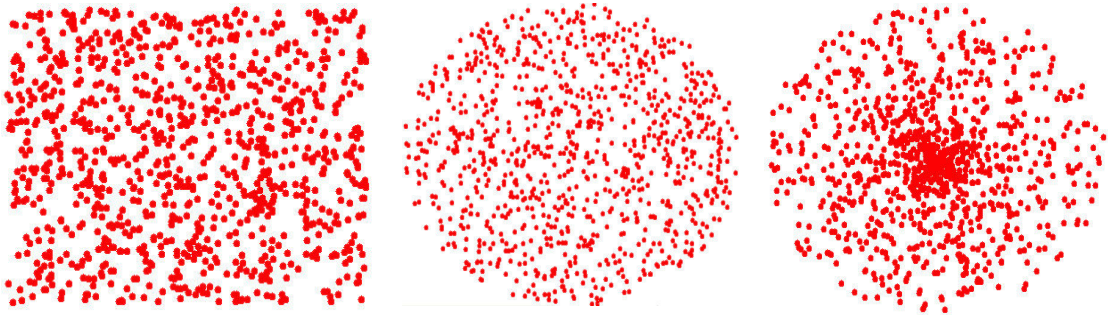

**Figure 5.** The three types of random data usually used to test progams

Now with the experimental result found in the previous paragraph random sets of data using one scheme or the other were tested for convex hulls. Results are shown in Figure 6 below.

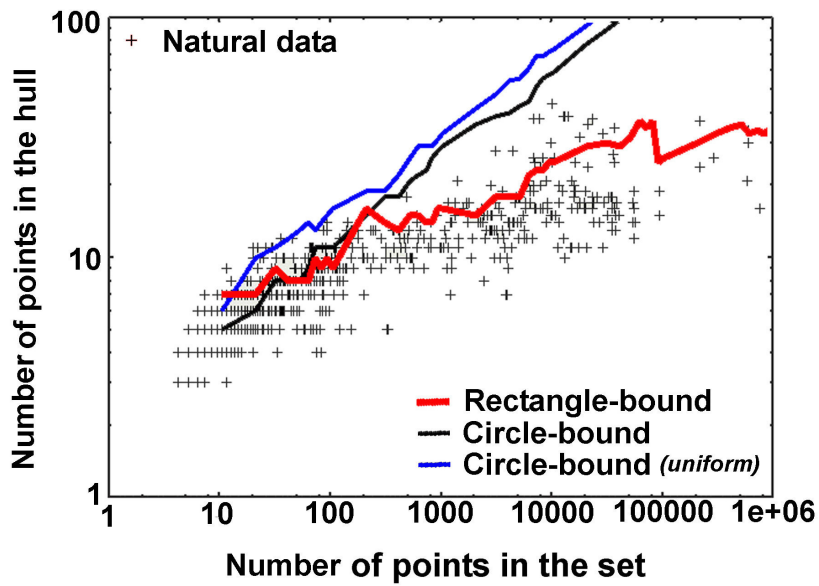

**Figure 6.** Number of convex vertices *vs* number of points in the set for random data

The obvious conclusion from this figure is that - at least from a convex hull computation point of view - random rectangle-bound datasets are much closer to natural data distributions than circle-bound datasets. This fact furthermore adds to the explanation of the special behaviour of the above-mentioned isolated case of the clusters datasets as it is also circle-bound.

## 5. Further work

Although some research and publications exist on the prediction of the number of convex vertices they are usually focused on polygons. One can cite for instance the work of Prekopa[8], Dumitrescu[5], Buchta[2,3] and most recently Nivash & al.[7]. Only mathematical studies on point sets orders or formal polygon decomposition are dealing with general datasets of points. The reader might refer to Arkin & al.[1] for further bibliography on this subject. However not being a mathematician the author will not discuss the implications of the results of this study in these fields and will let specialists deal with it.

## 6. Conclusion

This paper has presented some experimental results obtained on a large database of 790 datasets ranging from 4 up to 760,000 points of various origins. All the *natural* datasets as well as computer-generated ones avoiding worst-cases for the convex hull computation lead to an increase of the number of convex vertices with the number of input points which can be expressed as a range $2.35\ N^{0.091} \leq h \leq 19.19\ N^{0.091}$. This could be used as a basis for limiting memory usage in pre-allocated storage or start at a realistic value in some algorithms like Chan's, or to provide numerical bounds for output-sensitive algorithms. The large number of sets as well as their various origins strongly supports the truthfullness of the conclusion and its range. An additional conclusion is that random data built for the testing of convex hull algorithms should be rectangle-bound rather than circle-bound in order to represent reality.